\documentclass[twocolumn,prl,aps,floatfix,epsf,psfig,groupaddress,amssymb]{revtex4}
\usepackage{epsfig}
\usepackage{amsmath}
\pdfoutput=1

\begin{document}

\title{Spin-spin correlations of magnetic impurities in graphene}

\author{A. D. G\"u\c{c}l\"u}
\author{Nejat Bulut}
\affiliation{Department of Physics, Izmir Institute of Technology, IZTECH,
  TR35430, Izmir, Turkey}

\date{\today}

\begin{abstract}
We study the interaction between two magnetic adatom impurities in graphene using the
Anderson model.  The two-impurity Anderson Hamiltonian is solved
numerically by using the quantum Monte Carlo technique. We find that the
inter-impurity spin susceptibility is strongly enhanced at low temperatures,
significantly diverging from the well-known Ruderman-Kittel-Kasuya-Yoshida (RKKY) result
which decays as $R^{-3}$. 
\end{abstract}
\maketitle

Graphene\cite{Wallace47,NGP+09,Novoselov+Geim+04,Zhang+Tan+05},
a two-dimensional honeycomb lattice of carbon atoms, shows promise as
a material for nanoelectronics due to high electronic and thermal
conductivity. Moreover, graphene structures engineered at the
nanoscale are shown to give rise to unique magnetic properties due to
the formation of finite magnetic moments at the 
edges\cite{Wimmer+08,Yazyev+08,Yang+08,Jung+09,Rojas+09,Wunsch+08,Dutta+08,Guclu+09,Potasz+12,Guclu+13}, 
which could be important for nanoelectronic and spintronic device applications. Another way of
probing magnetism in graphene is through the exchange interaction between
impurity atoms mediated by the host electrons, known as RKKY
interaction\cite{power+13,brey+07,saremi+07,blackschaffer+10,sherafati+11,kogan+11}. 
Understanding the effective interaction between impurity
atoms in graphene is also important from fundamental physics point of
view since the excitations on a honeycomb lattice are
massless Dirac fermions, giving rise to a behavior different from
semiconductor or metal host structures\cite{power+13,brey+07,saremi+07,blackschaffer+10,sherafati+11,kogan+11}.

The RKKY interaction in graphene exhibits unique features different
from other two-dimensional systems. In Ref.\onlinecite{wunsch+06}, it was
predicted that RKKY interaction should decay as $R^{-3}$ in contrast
with $R^{-2}$ behaviour found in a two-dimensional electron gas\cite{fischer+75}, where
$R$ is the distance between the two impurities. This was later
confirmed in Ref.\onlinecite{saremi+07} where other important features
of RKKY interaction in graphene were clarified as well. In particular, a general
proof regarding the sign of the RKKY interaction  in a half-filled
bipartite lattice was given: interaction between moments sitting on
the same (opposite) sublattice(s) is ferromagnetic
(antiferromagnetic). We note that the biparticity of the graphene
lattice is also at the hearth of Lieb's theorem on magnetism\cite{Lieb+89} which
gives rise to edge magnetism in
graphene nanostructures\cite{Guclu+09,Potasz+12,Guclu+13}.  
In Ref.\onlinecite{saremi+07}, it was also shown that the RKKY
interaction is subject to an oscillatory term  of the type 
$1+cos(2\Delta{\bf k} \cdot {\bf R})$ ($2k_F$ oscillations), 
where $\Delta {\bf k}$ is the reciprocal lattice vector connecting
two Dirac points in the Brillouin zone. All these features were confirmed by
others\cite{blackschaffer+10,sherafati+11,kogan+11} using different approximation schemes. 
It should be
noted that impurities do not have to sit on top of a particular atom,
but can bond with  several neighboring atoms. In fact, mechanical and
electronic properties of various adatoms on graphene were previously investigated 
\cite{sevincli+08,zhou+09,xiao+09,jacob+10,wehling+10,chan+11,rudenko+12,saffarzadeh+12}. 
GGA+U calculations\cite{wehling+10}  show that the
particular bonding configuration may depend  on the value of the
on-site interaction parameter $U$. 
However, according to RKKY analysis \cite{saremi+07},
the interaction between plaquette-type impurities 
(where the impurity atom bonds with  the six
Carbon atoms of a hexagon in the honeycomb lattice) 
is much weaker and decays rapidly as $R^{-4}$. This is also consistent 
with our QMC results (not shown). Therefore, in the following we will focus
on the interaction between impurities with on-top configuration.

In this work, we use the Hirsch-Fye quantum Monte Carlo (QMC) 
method\cite{HirschFye+1986} to calculate the 
magnetic susceptibilities of the two-impurity Anderson model.
We find that, although the biparticity theorem of Ref.\onlinecite{saremi+07}
and the $2k_F$ oscillation behaviour are not  affected by electron-electron
interactions, the long range behaviour of the effective RKKY interaction
is strongly enhanced, becoming several orders of magnitude larger at longer
distances.

The two-impurity Anderson model for a graphene host is given by
\begin{eqnarray}
 \nonumber
H&=&\sum_{{\bf k}\alpha\sigma}\epsilon_{{\bf k}}c^\dagger_{{\bf k}\alpha\sigma}c_{{\bf k}\alpha\sigma}
  +E_{d}\sum_{i\sigma}d^\dagger_{i\sigma}d_{i\sigma}\\
&+&\sum_{{\bf k}\alpha i\sigma}\left( V_{{\bf k}\alpha i}c^\dagger_{{\bf k}\alpha\sigma}d_{i\sigma}+ \mbox{h.c.} \right)
 +U\sum_{i}n_{id\uparrow}n_{id\downarrow}
 \label{anderson}
\end{eqnarray}
where $c^\dagger_{{\bf k}\alpha\sigma}$ creates a host electron with
wavevector ${\bf k}$ and spin $\sigma$ in the valence $\alpha=v$ or
conduction $\alpha=c$ band,  $d^\dagger_{i\sigma}$ creates an electron
at the impurity site $i$, and
$n_{id\sigma}=d^\dagger_{i\sigma}d_{i\sigma}$. In addition, $U$ is the onsite
Coulomb repulsion and $E_{d}$ is the impurity energy level. The
electronic spectrum of the graphene host $\epsilon_{\bf k}$ and the
hybridization matrix elements $V_{{\bf k}\alpha i}$ are calculated
analytically in terms of graphene structure factor $f({\bf k})$ in the
nearest neighbour approximation with hopping parameter $t$. The
impurity-carbon atom hybridization parameter is denoted by
$V$. The calculations are performed within the symmetric Anderson model
where $E_d=-U/2$, as a function of inverse temperature $\beta=1/T$ and
the distance between the two impurities $R$.

The numerical results presented here were obtained using the
Hirsch-Fye quantum Monte Carlo technique which allows us to compute
the Matsubara single-particle Green's functions for impurity sites $i$
and $j$,
\begin{eqnarray}
   G^\sigma_{ij}(\tau)=-\langle T_\tau d_{i\sigma}(\tau) d^\dagger_{j\sigma}(0) \rangle,
\label{MGfunc}
\end{eqnarray}
where $T_\tau$ is the Matsubara time-ordering operator and $d_{i\sigma}(\tau)=e^{H\tau}d_{i\sigma}e^{-H\tau}$. 
In addition, we calculate the zero-frequency inter-impurity magnetic susceptibility using:
\begin{eqnarray}
   \chi_{12}(\omega=0)=\int_0^\beta d\tau \langle M^z_1(\tau)M^z_2(0) \rangle,
\label{staticchi}
\end{eqnarray}
where $M^z_i=n_{id\uparrow}-n_{id\downarrow}$. Local magnetic moment of impurity adatoms 
on graphene were studied in Ref.\onlinecite{hu+11}. Here we concentrate on the
impurity-impurity magnetic correlations.

\begin{figure}
\includegraphics[width=3.5in]{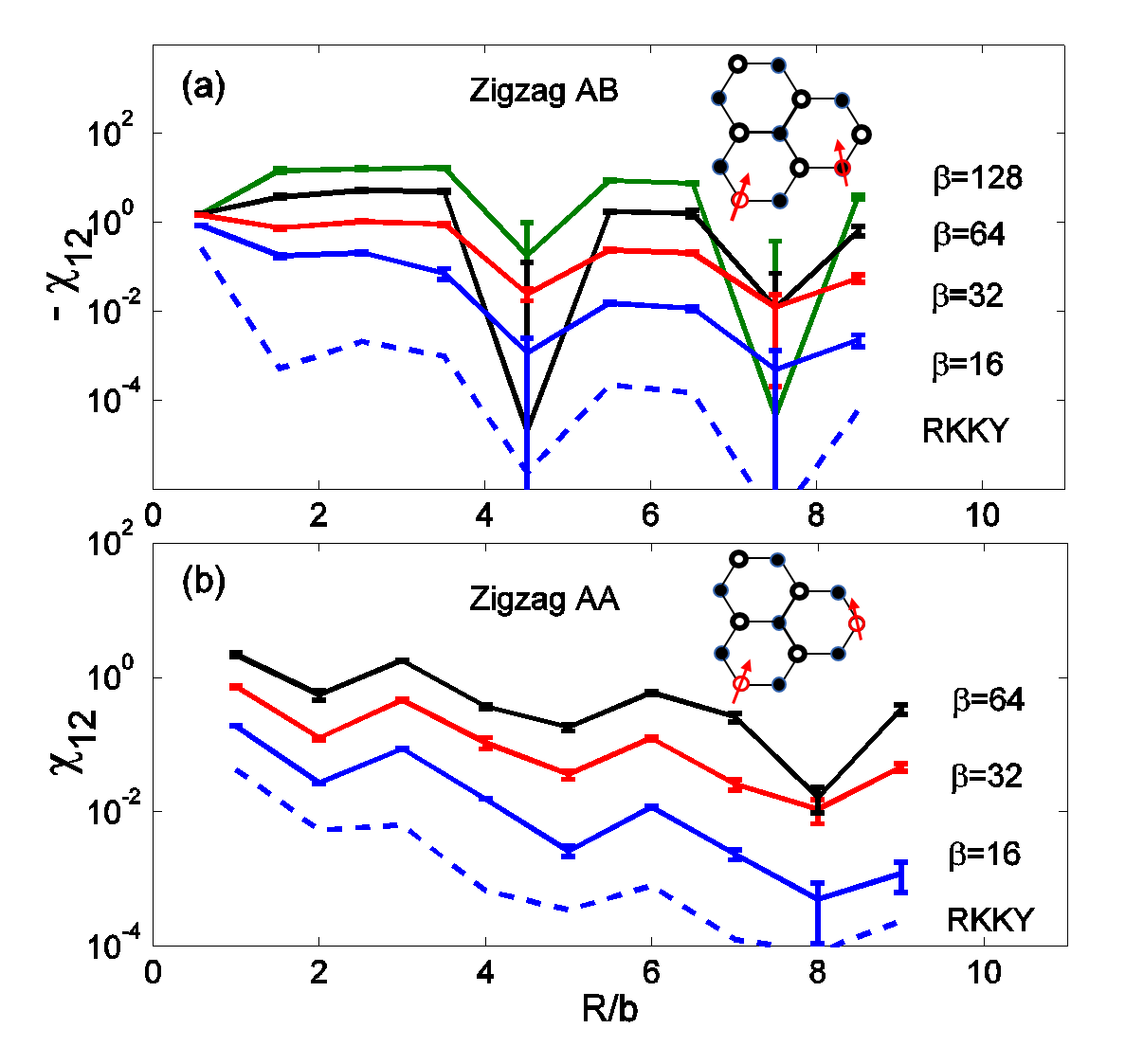}
\caption{\label{fig:1} The static magnetic susceptibility between 
two magnetic adatom impurities along the zigzag direction as a function
of distance for
(a) the AB configuration (impurities on opposite sublattices, shown in
the inset) and (b) the
AA configuration (impurities on the same sublattice, shown in the inset), 
obtained by
QMC calculations at different inverse temperatures $\beta$. The dashed
lines are the RKKY results from Ref.\onlinecite{sherafati+11}. The magnetic coupling 
obtained by the QMC
shows the same ferromagnetic and antiferromagnetic behaviour, 
and the $2k_F$
oscillations as in the RKKY results. However, at low temperatures, 
the effective magnetic coupling becomes much stronger and the QMC
results diverge from the RKKY's $R^{-3}$ decay. }
\end{figure}
In Fig.1, we consider the case where the two impurities are located
along the zigzag direction  of the honeycomb lattice, sitting on
different (zigzag AB, Fig.1a) and same (zigzag AA, Fig.1b)
sublattices. The static magnetic susceptibilities $\chi_{12}$ given in
Eq.\ref{staticchi} are calculated as a function of the distance
between the impurities $R$ (in units of the second nearest neighbour distance
$b$) at different inverse temperatures $\beta$ expressed in units of
$t^{-1}$. We take $V=t$ and $U=0.8t$ (see Fig.2 for larger values of
$U$). Here, the results are also compared to the analytical RKKY
results\cite{sherafati+11} donated by the dashed lines. For the AB configuration, the RKKY model yields
to an antiferromagnetic coupling between the two impurities as
seen from the sign of $\chi_{ij}$, and Fermi oscillations with
minima at every $(2+3n)${\it th} B-atom along the zigzag AB direction. For the AA configuration, 
the coupling is ferromagnetic and the oscillations have maximum
at every $(3+3n)${\it th} A-atom. For both cases, as already 
mentioned, the oscillations decay as $R^{-3}$. All these 
behaviours agree well with the Anderson model (QMC) results especially at
higher temperatures. However, the results are very sensitive to the
temperature. As the temperature is lowered,
significant deviations from RKKY results occur. The overall magnitude
of the static susceptibility increases by several orders especially at
larger $R$ values, the decay of RKKY becomes much slower, and the
Fermi oscillations become less prominent. Strikingly, at
$\beta=128t^{-1}$ which corresponds to $T=272$ K for $t=3$ eV,  there
is no decay in the range of $R$ studied here. 

\begin{figure}
\includegraphics[width=3.5in]{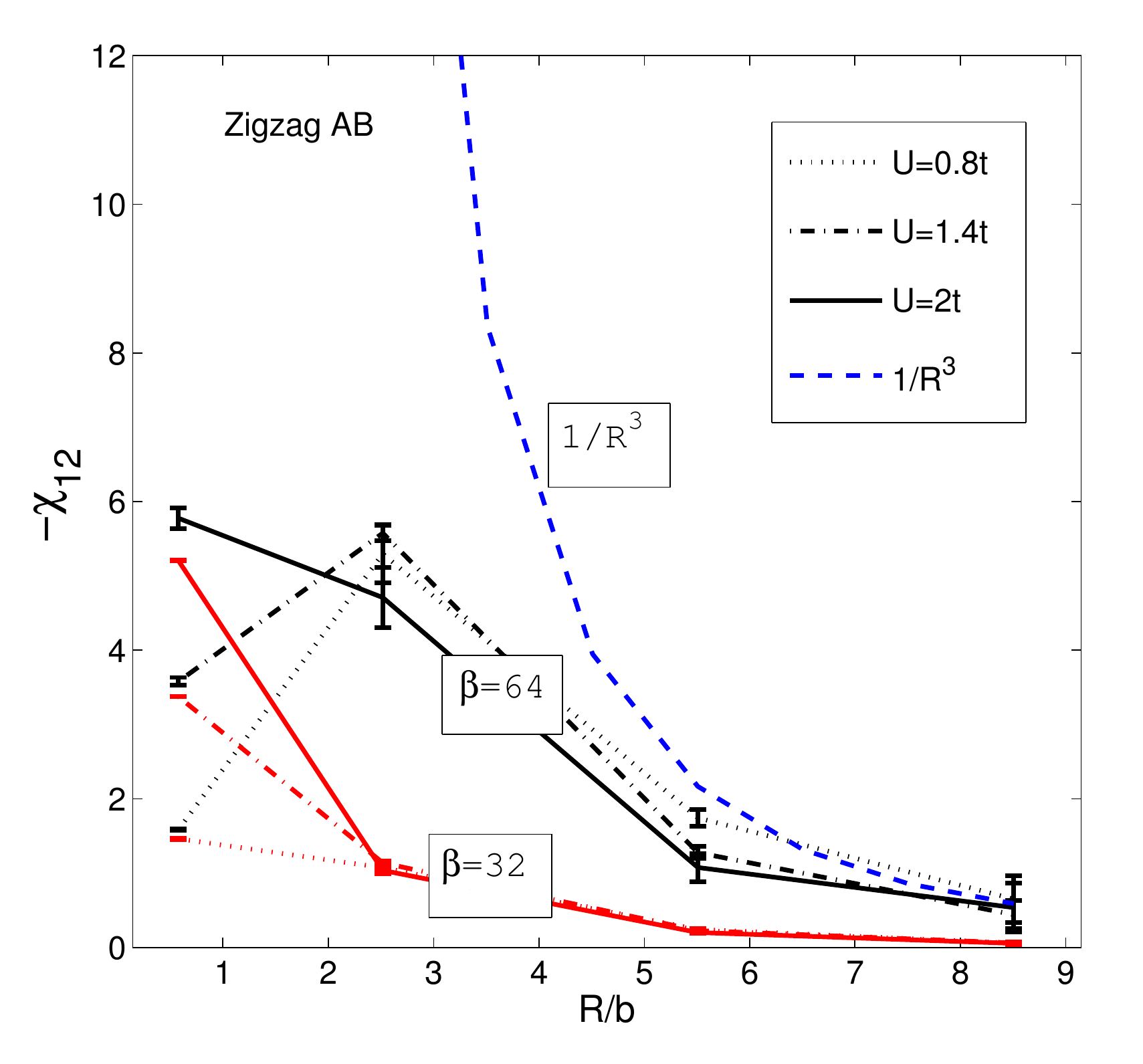}
\caption{\label{fig:2} The static magnetic susceptibilitiy between 
two magnetic adatom impurities along the zigzag direction as
a function of distance for
the AB configuration obtained using QMC method, at different 
inverse temperatures $\beta$ and different $U$. The dashed
lines shows the $R^{-3}$ decay predicted by the RKKY model. Even
at the highest $U$ value QMC calculations yield much longer-ranged
effective magnetic coupling between the adatoms.}
\end{figure}
In Fig.2, we investigate the effect of $U$ on the static
susceptibility. The susceptibilities are calculated at
$\beta=32t^{-1}$ and $\beta=64t^{-1}$ for the zigzag AB case (similar
to Fig.1a but in linear scale instead of logarithmic). Calculations
are repeated for $U=0.8t$, $1.4t$, and  $U=2t$, corresponding to 2.4
eV, 4.2 eV, and 6 eV, respectively. Although the exact value for $U$
is not known for 3d transition metal adatoms in graphene, its
effective value is estimated to be in the range of 2-4 eV
\cite{jacob+10,wehling+10,chan+11,rudenko+12}. As the statistical fluctuations increase for larger values
of $U$ in QMC calculations, the analysis is restricted
to four different values of $R$ corresponding to first, third, sixth, and
ninth atoms (along the zigzag direction) belonging to the maxima of
the RKKY oscillations. Clearly the main effect of increasing $U$ is to
increase the susceptibility for $R/b < 3$, i.e. at very short
ranges. For $R/b > 3$, we do not observe a significant change in 
$\chi_{ij}(R)$ within our statistical accuracy.
The overall behaviour thus becomes slightly closer in shape to the
$R^{-3}$ decay (dashed curve), but there are still several orders of
magnitude of difference between the RKKY and Anderson model results.
Thus, the main conclusions of Fig.1
remains unchanged for the values of $U$ considered here.

\begin{figure}
\includegraphics[width=3.5in]{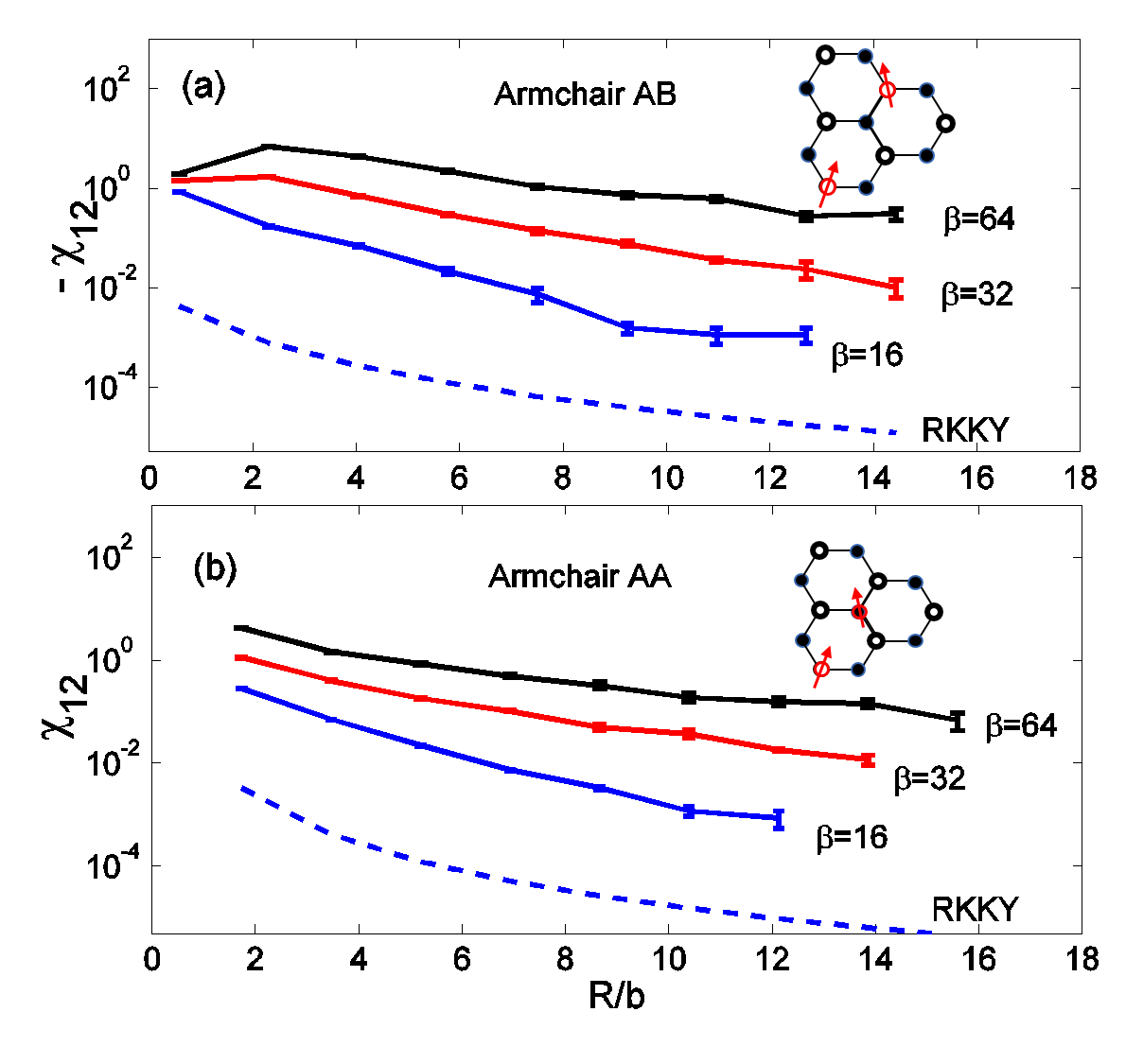}
\caption{\label{fig:3} The static magnetic susceptibilitiy between 
two magnetic adatom impurities along the armchair direction as a
function of distance for
(a) the AB configuration (shown in the inset) and (b) the
AA configuration (shown in the inset), obtained by
QMC calculations at different inverse temperatures $\beta$. The dashed
lines are RKKY results from Ref.\onlinecite{sherafati+11}. The magnetic coupling from QMC
show the same ferromagnetic and antiferromagnetic behaviour, 
At low temperatures, 
the effective magnetic coupling becomes much stronger and the QMC
results diverge from the RKKY's $R^{-3}$ decay. }
\end{figure}
We now turn to the armchair configuration. In Fig.3, the results
are presented for $U=0.8t$ at different values of $\beta$, for the
AB and AA configurations. Again the antiferromagnetic and ferromagnetic
phases for the AB and AA configurations are consistent with the RKKY
model. Note that along the armchair direction, the RKKY model does not exhibit
Fermi oscillations. This is also consistent with the QMC results at
higher temperatures (lower $\beta$) which show no clear structure within
our statistical accuracy. As the temperature is lowered, similar to the
zigzag case, the static susceptibility increases by more than two orders
of magnitude at larger distances of the order $R/b \sim 10$ 
significantly deviating from the $R^{-3}$ behaviour.

We note that the long-range behaviour of the impurity-impurity correlations observed 
in our numerical results for the Anderson model is consistent with
the predictions of Lieb's theorem for the Hubbard model 
in bipartite systems. According to Lieb's theorem\cite{Lieb+89},
if there is an imbalance between the number of A and B sublattice 
atoms, a finite magnetic moment $(N_A-N_B)/2$ arises at zero etmperature. In our case,
each impurity breaks the symmetry between the two sublattices
locally. Thus, if the impurities are far from each other, locally a 
finite magnetic moment must appear at each impurity location.  
If the two impurities are on same sublattices the magnetic moments must add, or  
otherwise cancel each other, giving rise to a strong ferromagnetic or antiferromagnetic
inter-impurity correlation.

\begin{figure}
\includegraphics[width=3.5in]{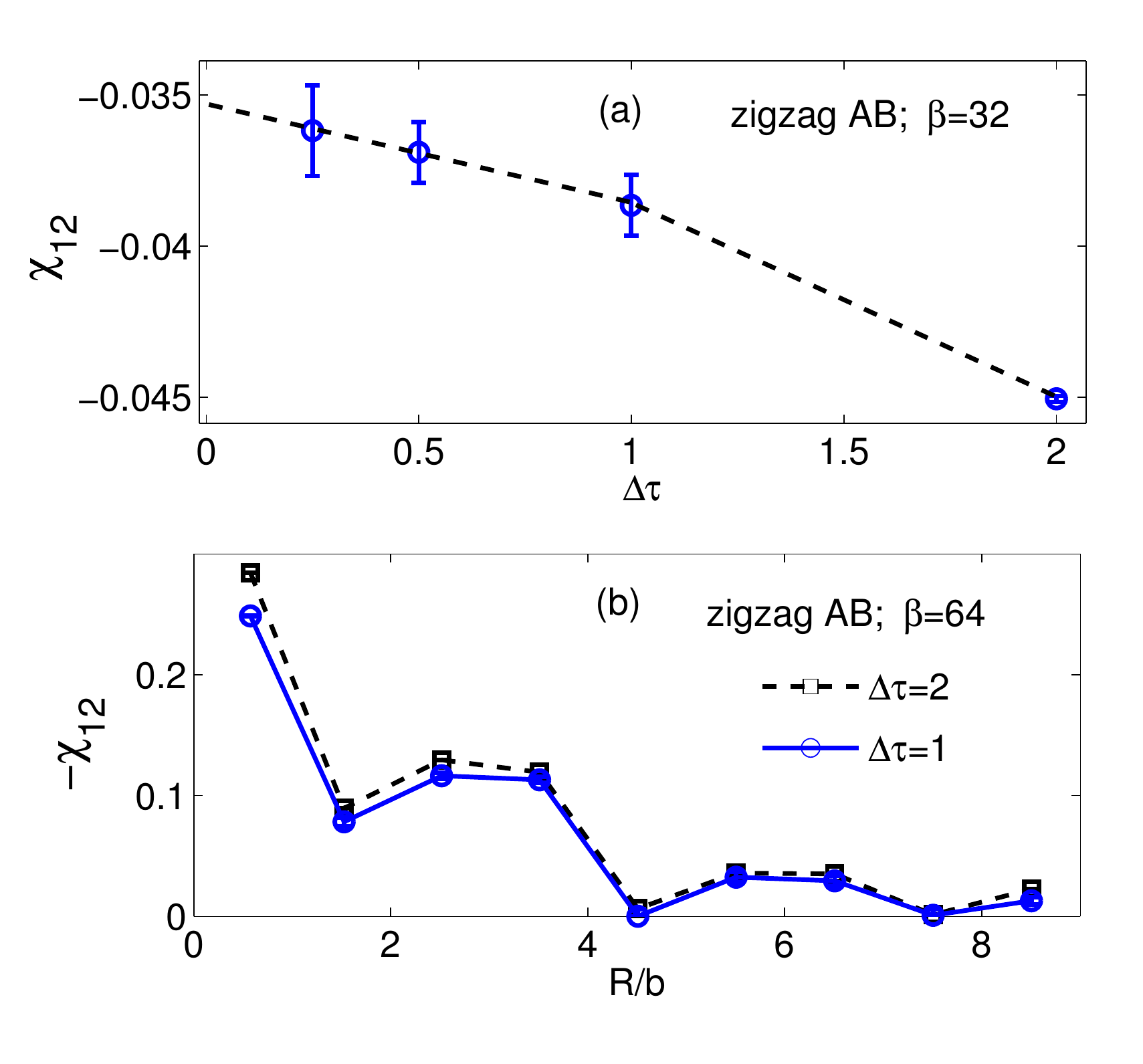}
\caption{\label{fig:4} The static magnetic susceptibilitiy between 
two magnetic adatom impurities along the zigzag for
the AB configuration obtained using QMC method, at different 
time-step $\Delta \tau$, (a) for $\beta=32t^{-1}$ and $R/b\sim 2.5$
(b) as a function of $R$ for $\beta=64t^{-1}$. These results show
that the finite time-step error is under control.  
}
\end{figure}
We now discuss the finite time-step error involved in numerical calculations. In
the QMC method, the partition function is discretized using 
$Z=\mbox{Tr}\prod^L exp(-\Delta \tau H)$ where $\Delta \tau$ is size of 
the time-step, $L$ is the number of Monte Carlo time-slices, and $\beta=L\Delta \tau$.
$Z$ defined above approaches the exact partition function of the system
in the limit of infinite $L$, i.e. small $\Delta \tau$. In order to check
the effect of using finite time-step, Fig.4a shows $\chi_{ij}(R)$ for the
third nearest AB-neighbours along the zigzag direction calculated for $\beta=64t^{-1}$
using $\Delta\tau = $ 2,
1, 0.5. and 0.25 in units of $t^{-1}$. Actual calculations are done for $\Delta\tau=1$
in previous figures. The estimated time-step error is within few error bars. We also
plotted in Fig.1b the results obtained for $\beta=64t^{-1}$ using $\Delta\tau = $ 2 and 1,
showing the finite time-step error is under control in our calculations

In conclusion, we studied the interaction between two magnetic adatom 
impurities in graphene within the Anderson model by using
the quantum Monte Carlo technique. Our results yield to the same magnetic
phases predicted by RKKY: ferromagnetic for the AA (same sublattice) 
configuration and antiferromagnetic for the AB (opposite sublattice) 
configuration. Moreover, $2k_F$ oscillations similar to those of RKKY exist. 
However, due to electron-electron interactions, the magnetic coupling between 
the impurities becomes more than two orders of magnitude stronger 
than what is predicted by the RKKY model, especially at lower temperatures.
In addition, the results significantly diverge from the $R^{-3}$ decay predicted
by RKKY and the effective interaction between the impurities
become long-ranged.


{\it Acknowledgment}. This research was supported by a BIDEP Grant
from T\"UBITAK, Turkey, and by a BAGEP grant from Bilim Akademisi -
The Science Academy, Turkey.



\end{document}